# "It took me almost 30 minutes to practice this".

## Performance and Production Practices in Dance Challenge Videos on TikTok

Daniel Klug
Carnegie Mellon University
dklug@cs.cmu.edu
http://daniel.klug.am

**Introduction**

Social media communication studies largely analyze audiovisual media online content regarding user processes of impression management and self-presentation. But little research looks at production practices in creating video artifacts to be shared on social media platforms. Video platforms, foremost YouTube, and videos on social media apps, such as Instagram, Snapchat, or Vine, have been studied regarding practices of consuming, creating, sharing, and engaging with audiovisual media artifacts (Cayari 2011, McRoberts et al. 2017, Phua et al. 2017, Vaterlaus et al. 2016, Yarosh et al. 2016, Yau & Reich 2019). However, the recent phenomenon of short video apps, and especially the highly popular TikTok, has not been sufficiently studied yet. TikTok, known as Douyin in China, as well as its predecessor musical.ly, is a music-based video sharing app. Compared to other services, TikTok content significantly involves users performing short sketches, memes, and especially dances based on musical or other audio snippets provided through the app. This paper uses the example of the #distancedance challenge on TikTok to analyze production practices from users' performances and paratextual information in 92 videos posted to the challenge. Based on theoretical implications of social practice, the paper asks:

> *RQ1: Who is participating in the dance challenge?*
>
> *RQ2: What are the audiovisual and textual characteristics of the submitted videos?*
>
> *RQ3: How do audiovisual and textual elements indicate users' production practices?*

The results show that #distancedance challenge videos are mostly done by white female teenagers in casual outfits filmed in a bedroom as sequence shot without video effects or text elements. Results of a qualitative content analysis illustrate users motivations and effort to learn the dance, and identify practices of adding individual performative elements, like gestures, to the dance performances. For further research, a profound analysis of video practices needs to be an integrated combination of product and production analysis.

**Theoretical Background and Related Work**

- *Social Practice*

Since the rise of social media, practices of amateur users creating, sharing, or engaging with (audio-)visual media artifacts have been analyzed from various social, cultural, political etc. perspectives. Various theoretical approaches, for example, uses and gratifications theory (Leung, 2009), involvement theory (Huang et al. 2010), or the technology acceptance model (Evans et al. 2014) explain why people create, share, or engage with photos and videos on social media. In the context of participatory culture (Jenkins 2019) and social media video (Burgess & Green 2018), this paper is situated within social practice theories (Giddens 2011 [1984], Schatzki, 1996) to analyze performance practices of users participating in a TikTok dance challenge regarding the presentation of the self (Goffman 2002 [1959]).

As a social practice, performance refers to the subject and the physicality needed to act and enact a practice. For example, an individual has to use their arms etc. to perform a practice in order for the organized and planned components of a practice to manifest. Practices are dynamic, they are constituted by language and actions and therefore connected to physical behavior (Schatzki 1996). Following Giddens (2011 [1984]), someone creating and sharing content online can be seen as an agent who reflects on their actions and is able to explain reasons for their actions and performances. Actions and performances to create online content can be analyzed through this inductive perspective of creators as agents and users as "video authors" (Yarosh et al. 2016).

- *Participatory Video Culture and Practices*

Research largely focuses on participatory practices of young(er) users and audiences (Ito et al. 2019, Lange 2014). For YouTube, foremost Burgess argues, in the tradition of Jenkins' (Jenkins 2009), that any video has cultural value as it can trigger creativity in many other platform users and that videos are "mediating mechanisms via which cultural practices are originated, adopted and (sometimes) retained within social networks" (Burgess 2014).

Due to its newness, little research can be found on the specifics of TikTok, studies mostly look at technical aspects of the app and its use (Chen et al. 2019, Lu & Lu 2019, Zhang, Wu & Liu 2019). Khattab (2019) analyzes users' facial expressions and attire in three TikTok challenges regarding body representation and finds the app can be a platform for "understanding the role of the body image in shaping notions of beauty and gender" (Khattab 2019).

Research on video production practices of youth authors largely concerns visual communication related to impression management online (Manovich 2009). Studies analyze, for example, social functions of selfies as gendered practices (Burns 2015), differences in shared content between YouTube and Vine (Yarosh et al. 2016), presentation and enunciation practices of Spanish YouTubers (Scolari & Fraticelli 2019), or user strategies of choosing what artifacts and narratives to share on Snapchat (McRoberts et al. 2017). Guerreo-Pico et al. (2019) identify casual, aspirational, and expert produsers to demonstrate how people acquire video production skills, who they share videos with, and how video production relates to modes of self-expression. However, these studies primarily look at how young(er) users select self-created media artifacts to manage social media representations of the self. Studies rarely look at motivations, practices, and strategies of users producing media artifacts that are intended to be shared in social media communities. For TikTok and its predecessor musical.ly, research on social practices in video production needs to consider that these are music-based short video apps. This implies sound- and music-related performative actions, like lip syncing or dancing, in a much stronger way than for YouTube, Snapchat, Instagram or the discontinued Vine. The characteristics of TikTok as a highly popular app and motivations, video creation practices, and consumption strategies of its users have not been researched yet.

**Characteristics of the Short Video App TikTok**

As its own format, short-form videos were first introduced in 2013 by Twitter's since discontinued Vine app. While the six to seven long Vine videos corresponded with the at the time shortness of tweets, they introduced features such as automatic video loops or hashtags to the presentation of short-form videos (Vandersmissen et al. 2014, Zhang et al. 2014) that were adopted by subsequent short video apps like musical.ly, Lasso, Bilibili, Byte (the unofficial Vine successor), and foremost TikTok.

    TikTok[1], in China known as Douyin, is a short video creation and sharing app owned by Chinese company ByteDance. Released in September 2016, it merged with musical.ly, a similar

---

[1] https://www.tiktok.com [31.3.2020].

short video app, in August 2018. Since then, TikTok rose to be the number one downloaded app in the App Store, and number three on Google Play[2] in the beginning of 2019. In December 2019, TikTok was downloaded 4.6 million times in the U.S.[3] and had a 40% daily engagement rate[4]; 37.2 % of U.S. TikTok users are teenagers[5]. As of January 2020, TikTok had around 800 million users[6], with its most monthly downloads ever (113 million)[7] in February 2020 most likely favored by COVID-19 related social distancing and people staying home[8].

In TikTok, users can create and post 15 seconds to one minute long videos and share them with the community. Videos can be viewed by anyone with an account or access to the direct video link. In contrast to similar apps, such as Vine, TikTok allows users to add features like stickers, text, or visual effects to videos (Lu & Lu 2019). Video creation is largely based on musical snippets, often popular songs, and users lip syncing the lyrics and performing to the music. By adding matching hashtags, users create collections of thousands of videos related to a song snippet or sound (Anderson 2020). A main practice of TikTok users is to create a performance that, through contextual knowledge, transforms a line of lyrics into a new statement, meme, or viral phenomenon. This is usually done by adding text elements and emojis in the video or hashtags in the video caption to create the targeted context. Creating TikTok videos often involves using filters and features offered by the app. For example, the speed manipulation feature allows users to slow down the music track or speed up the video recording to better sync moves, gestures, and lip syncing to the music and to precisely edit the video (Bresnick 2019). In general, TikTok features and design encourage users to create, remix, and join so-called challenges by using hashtags that are associated with challenges (Bresnick 2019). The app's characteristic feature is "duets", in which users can create a side-by-side split screen video with the original and react to it, almost in a form of call and response (Bresnick 2019).

---

[2] https://sensortower.com/blog/top-apps-worldwide-q1-2019-downloads [31.3.2020].
[3] https://www.statista.com/statistics/1090587/tiktok-ios-downloads-country [31.3.2020].
[4] https://www.statista.com/statistics/290492/mobile-media-apps-daily-engagement-rate-of-us-users [31.3.2020].
[5] https://www.statista.com/statistics/1095186/tiktok-us-users-age [31.3.2020].
[6] https://datareportal.com/social-media-users?rq=tiktok [4.4.2020].
[7] https://sensortower.com/blog/tiktok-record-revenue-downloads-february-2020 [31.3.2020].
[8] https://www.hitc.com/en-gb/2020/03/30/tiktok-why-you-so-negative-song-challenge-revealed [31.3.2020].

In the app, TikTok videos are presented one at a time in a loop until the user swipes up to the next one or pauses a video. Videos appear as a flow of content that users scroll through not knowing what the next video will be as there is no choice to intentionally select content, besides on a users' profile page (Anderson 2020). In this way, TikTok is closer to Vine than to Instagram stories, but it is also more playful than it is social. TikTok seems more like an experimental audiovisual playground for users rather than a social network to connect through (Anderson 2020, Bresnick 2019).

**The #distancedance Challenge on TikTok**

Dance challenges are one of the most popular content on TikTok which is no surprise given the fact that TikTok videos are generally based on short music files and spur the practice to remix and reuse music (Bresnik 2019). Dances, like The Renegade[9] or Spooky Scary Skeleton[10], are often initiated or made popular by TikTok influencers, for example, Hype House members Chase Hudson, Avani Gregg, Nick Austin, Addison Rae, or Charli D'Amelio and her sister Dixie[11]. These dances are subsequently imitated, recorded, and posted by regular users motivated to participate in these challenges. Besides learning and showing off one's dance skills, challenges also create user communities around creative amateur peer creation and participation in social online environments.

---

[9] https://www.cnn.com/2020/02/16/us/renegade-dance-tiktok-k-camp-trnd/index.html [31.3.2020]
[10] https://www.rollingstone.com/culture/culture-features/spooky-scary-skeletons-tiktok-meme-895887 [31.3.2020].
[11] https://www.nytimes.com/2020/01/03/style/hype-house-los-angeles-tik-tok.html [31.3.2020].

On TikTok, various viral challenges such as #FlipTheSwitch[12] (people switching outfits in front of a mirror) or #wineglasschallenge[13] (leaning back and trying to pour the wine into the mouth of a person sitting behind you) receive particular popularity during the COVID-19 pandemic as they keep people entertained during social isolation at home[14]. While such viral challenges show how people are using humor to cope with their quarantine situation[15], dance challenges as well highlight semi-professional creative outputs in digital media environments[16]. Studies and interviews with TikTok users show that especially younger people are motivated to participate in challenges if they address existing individual skills or create a personal motivation to learn skills, for example, dance moves (Ahlse, Nilsson, Sandström 2020). Research on the viral "Ice Bucket Challenge" on Facebook shows people attested a higher social capital because they are, for example, more extroverted and open to new experiences are more likely to participate in viral challenges (McGloin, Oeldorf-Hirsch 2018). While this paper does not make any claims regarding motivations to participate in TikTok challenges, it can be assumed that the open, performative, and little structured entertainment and pastime centered TikTok environment presents a lower social capital barrier for users to participate in TikTok challenges. Yet, underlying factors such as narcissism, attention-seeking, and wanting to create a positive self-presentation online (Bergman et al. 2011) likely apply to participation in TikTok dance challenges as well.

---

[12] https://www.tiktok.com/tag/fliptheswitch [2.4.2020].
[13] https://www.tiktok.com/tag/wineglasschallenge [2.4.2020].
[14] https://www.nytimes.com/2020/03/26/style/viral-challenges-coronavirus.html [2.4.2020].
[15] https://www.vox.com/2020/3/6/21167987/world-humor-curb-coronavirus-spread-washington-singapore-vietnam [2.4.2020].
[16] https://www.latimes.com/projects/la-social-media-dance-influencer/ [2.4.2020].

Fig. 1: Screenshot of the video by Charli D'Amelio with used paratext (red box)

On March 24, 2020, Charli D'Amelio[17], the biggest star on TikTok, posted a video[18] tagged with #distancedance. D'Amelio, a trained dancer famous for her TikTok dance videos, initiated this dance challenge in cooperation with sponsor *Procter & Gamble* to raise money for *Feeding America* and *Matthew 25*, two organizations that help in-risk populations during the 2020 COVID-19 pandemic[19]. Therefore, the #distancedance challenge contributes to TikTok

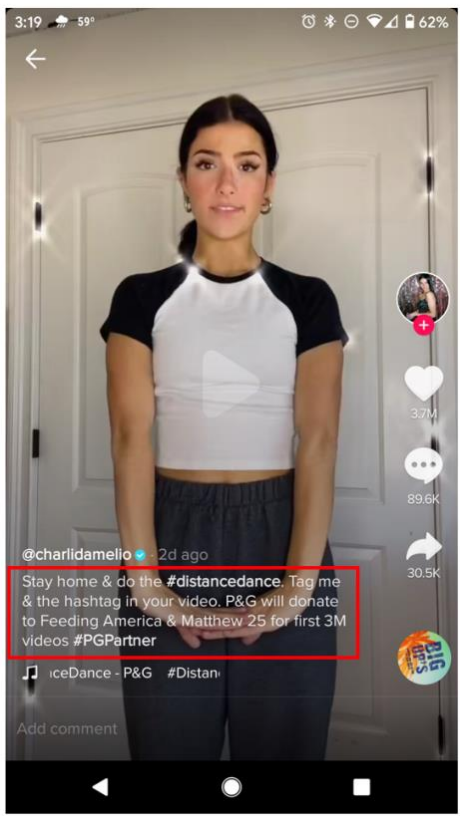

dance trends, but tries to create awareness by its name referring to the order of social distancing to prevent the spread of the virus. The video caption explains *P&G* will donate money for the first three million videos created doing the dance, therefore, users participating should use #distancedance and add Charli D'Amelio (@charlidamelio) in their videos. As of April 6, 2020, #distancedance had 8.8 billion views, Charli D'Amelio's video had 187.1 million views and 5.8 million likes. The #distancedance challenge is set to a part of the song "Big Ups" by Jordyn and Nic Da Kid Ft. Yung Nnelg[20]. "Big

---

[17] https://www.tiktok.com/@charlidamelio [31.3.2020].
[18] https://www.tiktok.com/@charlidamelio/video/6807971434959310085 [27.3.2020].
[19] https://www.insider.com/charli-damelio-distance-dance-on-tik-tok-to-drive-donations-2020-3 [31.3.2020].
[20] https://www.youtube.com/watch?v=42E9UcYkVMQ [31.3.2020].

Ups" is a mid-tempo dance song, and it can be assumed the distinct part was chosen because of the lyrics. In the context of the #distancedance challenge, the starting line of the song segment ("Why you so negative") appears to encourage users not to be depressed by the circumstances of the COVID-19 crisis, and rather help by participating in the sponsored challenge. The end of the song segment ("inhale, exhalte, breathe slow, rewind") is clearly instrumented to tell viewers and participants to stay calm and relax during the crisis.

Charli D'Amelio's video shows her performing the dance moves in front of a white door, most likely a closet in her bedroom. The video uses the TikTok "bling" effect, she wears blue-grey long pants and a white-black T-shirt. Her hair is tied back in a braid, she is wearing earrings. Charli lip syncs the first line of the song lyrics ("why you so negative"), after that she gestures and mimes with mouth and eyes in sync with the dance moves. At the end, she walks off screen making a "V sign" gesture and adds the text element "inhale, exhale, breathe slow, rewind. stay at home!💕" . The first part of the text quotes the song lyrics heard right before the end while "stay at home!" appeals to users to follow government orders of self isolating during the COVID-19 crisis. The 💕 emoji indicates Charli's empathy, love, and encouragement for her fans and the participants of the challenge.

**Data Collection and Methodological Approach**

On March 27, 2020, the first 100 TikTok videos listed that day under #distancedance on the TikTok website[21] were collected[22]. Each video was saved and a screenshot was taken to document profile information, used hashtags (#), tags (@), and video captions. Distribution or popularity of profiles and video (followers, likes, shares) were not relevant for this study. The

---

21 https://www.tiktok.com/tag/distancedance?lang=en [27.03.2020].
22 For anonymity reasons no direct links to videos are included in the paper.

eight earliest dated videos in the sample were not based on the song "Big Ups" and did not show dance performances. Therefore, as they used #distancedance in a different context, they were excluded resulting in a final sample of 92 videos. First, the 92 videos were analyzed regarding their visual content, that is, who is acting and what elements can be seen, and regarding paratextual elements (see Fig.1) (Genette 1997), that is, what hashtags, tags, and caption was added to the description of the video (Simonsen 2014).  This was done using a qualitative content analysis approach (Flick 2014) for audiovisual media artifacts (Altheide & Schneider 2012). The visual content of each TikTok video was coded by the following audiovisual and performance related categories:

| *Analytical category (video)* | *Guiding question related to video content* |
| --- | --- |
| Type of video | Which cinematic style was the video filmed or edited in? |
| Video effects | Which visual effect provided by the app does the video use? |
| Text elements | Which text elements are added to the video? |
| Lip syncing | Does the user in the video lip sync the song lyrics? |
| Number of people | How many people perform the dance in the video? |
| Gender | What is the gender of the people performing in the video? |
| Ethnicity | What is the ethnicity of people performing in the video? |
| Age range | What is the rough age range of the people performing in the video? |
| Outfit | What outfit are people wearing in the video? |
| Setting | Where do people perform the dance in the video? |

In addition, each video paratext was coded by the following text related categories:

| *Analytical category (paratext)* | *Guiding question related to paratextual content* |
| --- | --- |
| Number of hashtags (#) other than #distancedance | How many hashtags were used in the video description in addition to #distancedance? |
| Number of tagged users (@) other than @charlidamelio | How many users were tagged in the video description other than Charli D'Amelio? |

| | Caption | Does the video have a caption in the description (other than hashtags/tags)? |
|---|---|---|

The sampled videos were then analyzed regarding visual and (para-)textual indicators that allow to draw conclusions on users' video creation strategies and performance practices in participating in the #distancedance challenge. In comparison to Charli's original, the videos were openly coded along the following categories:

| *Analytical category (strategies and practices)* | *Guiding question related to video creation strategies and performance practices* |
|---|---|
| Textual | Does the video (para-)text indicate information about creating the video or performing the dance? |
| Performance | Does the video include performance elements that indicate performance practices? |
| Appearance | Does the visual appearance of users in the video indicate strategies of creating the video? |
| Setting | Does the setting of the performance in the video indicate strategies of creating the video? |

**Results**

- *Demographics of the #distancedance Challenge Participants*

The study can not make any representative claims regarding correct age, gender, or ethnicity of the people seen in the sampled videos as these categories are very contested and pose a great risk of misgendering and misracing people. However, it seemed that some assumptions can be made for age and gender based on common visual indicators, profile information, and video descriptions. This only serves to describe general impressions of who is participating in this TikTok dance challenge without relevance for further analysis and interpretation.

According to these observations, #distancedance videos are mainly performed by a single white female teenager in a casual outfit in their bedroom. Based on visual indicators in the video content, profile information, and paratextual video information, 64 users (69.6%) were identified as female, 27 (29.3%) as male, and one person was identified as non-binary. 53 participants were categorized as teenagers (13 to 19 yo) (57.6%), 34 as twentysomethings (20 to 29 yo) (37%), three as over 30 years old, and two users' age was not defined. The latter concerns a video of a woman wearing a face mask and a video by Brutus Buckeye, the athletes mascot of Ohio State University, who is described as male[23], yet has no age or ethnicity which is why they were coded as undefined.

| no. of people | | gender | | age | |
|---|---|---|---|---|---|
| one person | 83 | female | 64 | teenager | 53 |
| two persons | 7 | male | 27 | twentysomething | 34 |
| three persons | 2 | non-binary | 1 | over 30 | 3 |
| | | | | undefined | 2 |

Table 1: Demographics of people participating in the sampled #distancedance challenge videos (n= 92)

- *Artifact Characteristics of #distancedance Challenge Videos*

Out of the 92 videos, 86 were filmed as a sequence shot (93.5%), meaning they consist of one long take without cuts or editing (Bordwell, Thompson, Smith 2020). Only five videos (5.4%) were TikTok duets[24], and one was a montage of multiple scenes. 70 videos (76.1%) in the sample did not use any filter or effects, 18 videos (19.8%) used the bling effect like Charli in her video, four videos (4.3%) used a different effect or multiple effects. The large majority (83 videos, 90.2%) were solo performances (including the "duet" videos), only seven videos (7.6%) had two people performing, two videos (2.2%) had three people. As dance moves take time to learn and coordinate, it seems reasonable that users mostly participate alone or occasionally as a couple.

---

[23] https://en.wikipedia.org/wiki/Brutus_Buckeye [31.3.2020].
[24] A TikTok duet is a video in which the video author takes an initial video by another user and reacts or replies to it in a split screen mode which means that a TikTok duet consists of these two videos side by side.

In 28 videos (30.2%), users performed lip syncing the song lyrics or large parts of it. Lip syncing requires the user to know the lyrics and to have the physical and coordinative ability to mime the lyrics while also dancing. The appeal of TikTok clearly roots in using lip syncing to recontextualize song lyrics and other quotes, yet, dance challenges focus on the dance performance. This might explain a lower percentage of lip syncing videos; however, non-existing lip syncing also indicates that users focus their practice on the dancing and maybe did not put more effort into learning lip syncing in favor of posting their video as soon as they are pleased with their dance performance.

| type of video | | video effects | | text elements | | lip syncing | |
|---|---|---|---|---|---|---|---|
| sequence shot | 86 | none | 70 | yes | 17 | yes | 28 |
| duet | 5 | bling | 18 | no | 75 | no | 64 |
| multiple scences | 1 | other/multiple | 4 | | | | |

Table 2: Characteristics of the sampled videos (n = 92) submitted to the #distancedance challenge

17 videos (18.5%) added text elements to the video. Five of those were TikTok duets that added the complete text at the exact time and in the style of Charli's video. The other twelve users mostly used only parts of the original text element, occasionally adding words or making variations. For example, some users only added "stay at home" or "stay home & stay safe" at the beginning or throughout the video, or lyrics when they appear in song at the end of their videos or the beginning ("why you so negative"). Instead of the original text elements, some users added "distant dance" or "#distancedance" as another variant referring to the cause of this dance challenge. A few users added different text elements, such as longer, mostly motivational or empathetic statements referring to staying home, dealing with isolation and quarantine during the COVID-19 crisis, or commenting in a self-reflexive way on their performance of the dance. One video by a user who makes TikTok dance tutorials featured textual descriptions of the dance moves.

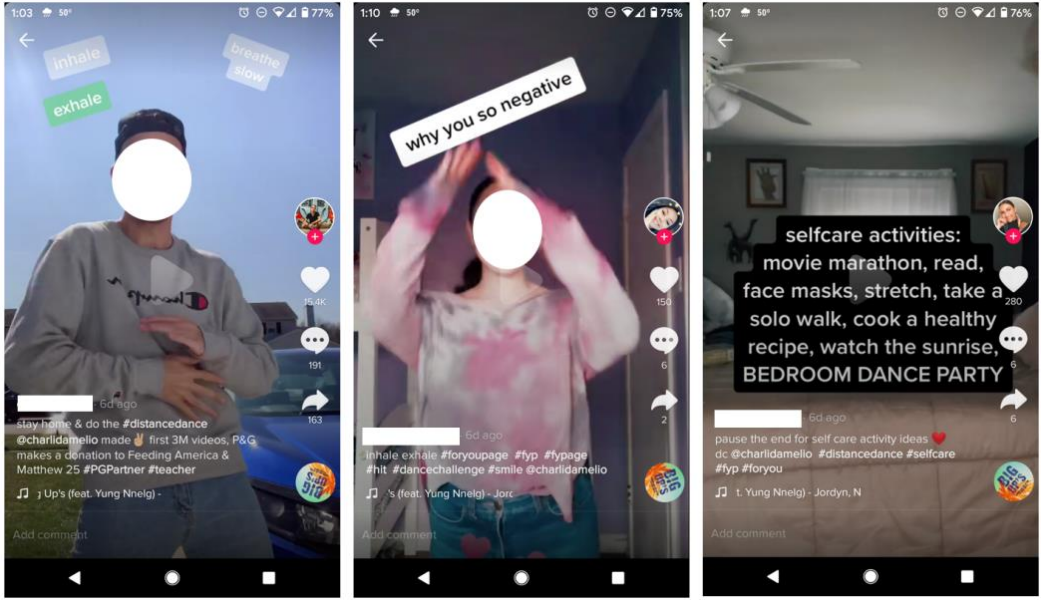

Fig. 2: Examples of different text elements used in videos

- *Outfits of Participants and Performance Settings in #distancedance Challenge Videos*

Regarding style and setting, users performed the #distancedance mostly in their bedroom wearing casual everyday outfits. In 65 videos (70.7%), the outfit was coded as casual, in eleven videos (12%) male and female users alike wore significantly short clothes, for example, tank tops or shorts. In five videos (5.4%), users were coded as elegantly dressed, meaning they wore nicer than casual clothes, and eventually makeup or accessories. Another five videos showed users performing in sporty or sport related clothes, like jerseys or fitness wear. Four users (4.3%) performed in pyjamas in a bedroom, two videos, one person wearing a dentist uniform and the mascot Brutus Buckeye, were labeled "other". Choosing an outfit for filming oneself seems like a deliberate decision, which indicates few users dressed up in a possible act of planning their dance performance with a certain outfit.

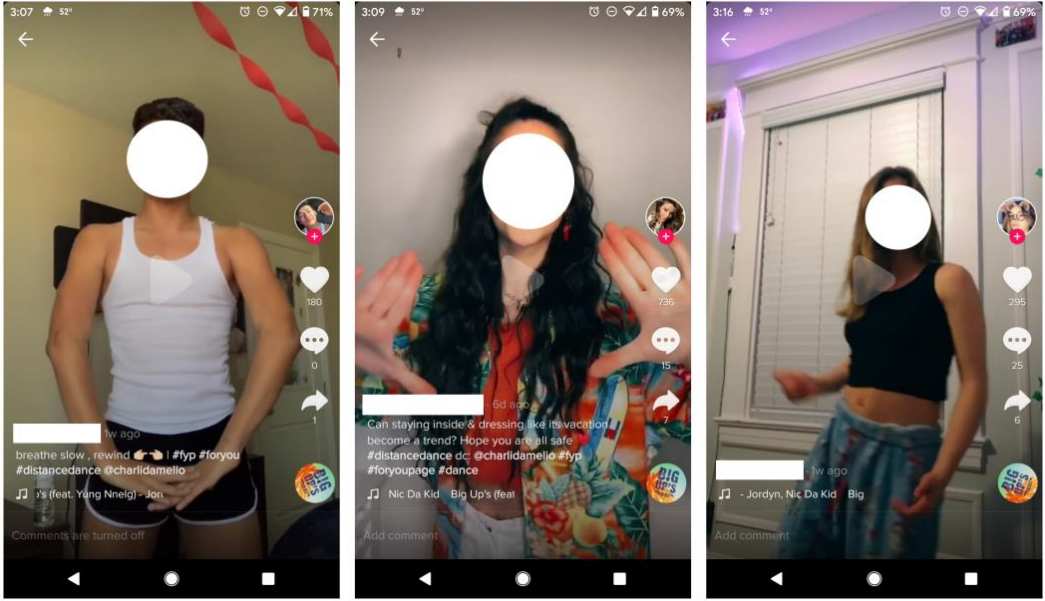

Fig. 3: Example of users' outfits in the sampled videos (left to right: short, elegant, pyjama)

While this may also be true for casual outfits, everyday clothes can be connected to everyday life, meaning the #distancedance performance is a less prominent part of the users' life and rather a common social activity that can be done at any time a day without needing specific circumstances or settings. The fact that some users wear pyjamas might back up that hypothesis meaning it could be a common activity to make TikTok videos before going to bed or after waking up.

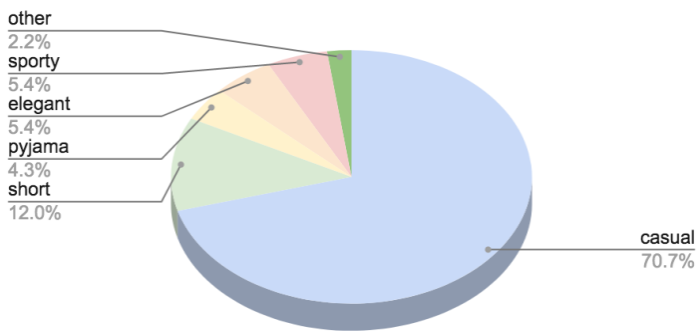

Fig. 4: Percentage of users' outfits in the sampled videos (n = 92)

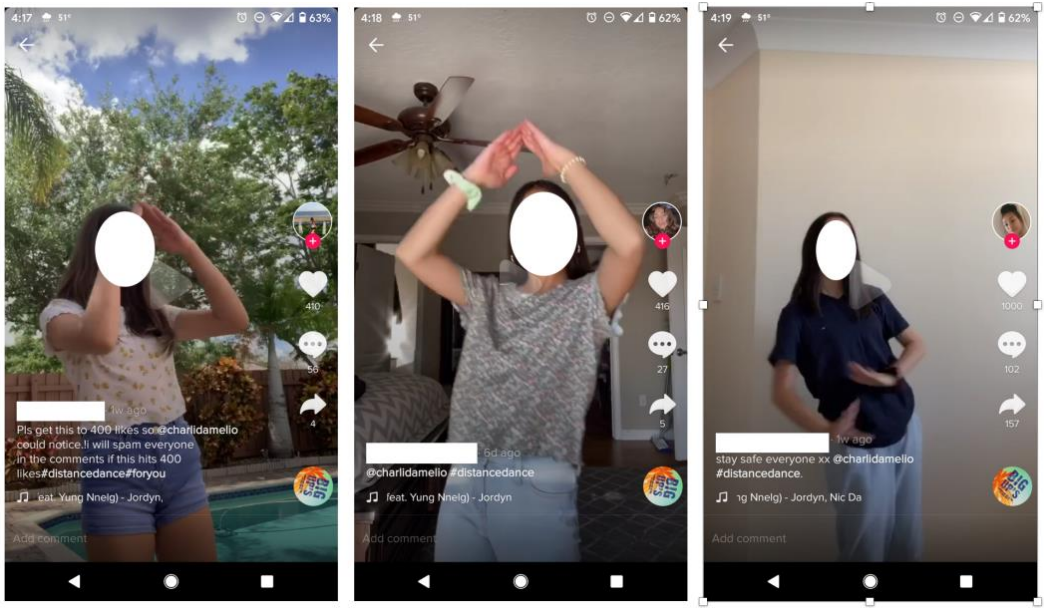

Fig. 5: Examples of setting of #distancedance videos (left to right: backyard, bedroom, in front of wall)

In this context, the settings in which users performed the #distantdance correlates with the assumption of more or less spontaneous TikTok performances as part of users' everyday life. 33 videos (35.9%) are set in a bedroom, 16 videos (17.4%) in a living room, and eleven (12%) in a hallway. All of these settings provide enough space, eventually privacy, and mostly good enough lighting for performing and recording the dance. Choosing a setting is an act of planning; however, only few of these videos include other spatial or decorative objects or elements that would indicate more thought put into planning. Yet, the variety of settings as such could indicate strategic choices of settings. For example, performing in front of a plain wall, inside or outside, eliminates distractions and allows to focus on the performance and the performer, while a backyard or porch may add more appealing visual elements to it. Kitchens and bathrooms might provide better lighting but less space, the latter allows users to perform in front of a mirror (or with enhanced privacy). As for settings, the 92 videos in the sample also included single occurrences of a gym, a basement, a mirror wall, and, in case of the only multiscene video, multiple settings.

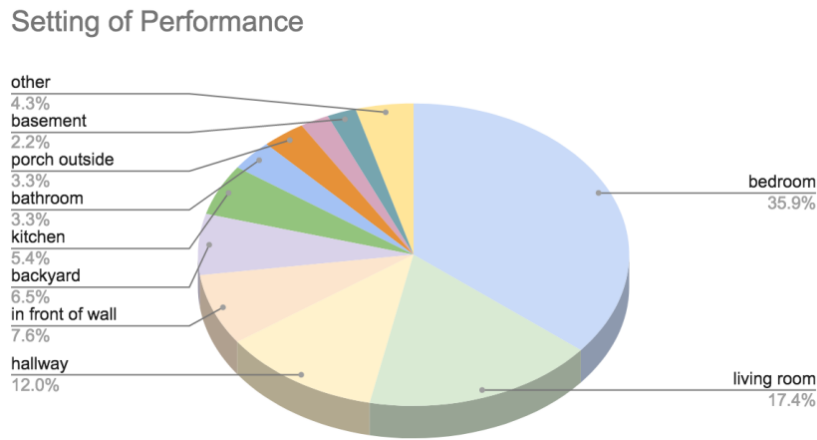

Fig. 6: Percentages of settings in which users performed the #distancedance in the sampled videos (n = 92)

- *Paratextual Characteristics of #distancedance Challenge Videos*

Looking at the paratextual elements, the majority of videos included the hashtag #distancedance, tagged @charlidamelio, and also had some kind of caption. Only three videos in the sample (3.3%) did not use any hashtag, five videos (5.4%) used a hashtag other than #distancedance, 20 videos (21.7%) only used the hashtag #distancedance, and the large majority of 64 videos (69.6%) used the hashtag #distantdance in combination with other hashtags. In total, the sample videos used a total of 348 hashtags with an average of 3.9 hashtags per video. The maximum number of hashtags added to a video was twelve, the maximum number of tagged users was two.

| no. of hashtags (#) added to video | | no. of tags (@) added to video | | video captions | |
|---|---|---|---|---|---|
| #distantdance and others | 64 | tagged Charli D'Amelio | 65 | yes | 65 |
| only #distantdance | 20 | tagged no one | 19 | no | 27 |
| only other hashtags | 5 | tagged someone else | 8 | | |
| no hashtags | 3 | | | | |

Table 3: Number of videos posted to #distancedance (n = 92) that used paratextual elements

65 users (70.7%) tagged only Charli D'Amelio (@charlidamelio) or Charli and other users, 19 users (20.7%) tagged no users, and 8 users (8.7%) only tagged someone else than Charli. 65 videos (70.7%) included a caption, 27 videos (29.3%) did not.

The results show that users participating in the #distancedance challenge mostly followed the challenge guidelines by using the hashtag and tagging Charli. However, almost a third of users did not and were only associated with the #distancedance challenge through the song file of "Big Ups" which in TikTok is linked to the challenge.

- *Hashtags Used with #distancedance Challenge Videos*

Besides #distancedance, videos also featured hashtags like #trending, #trend, #viral #viralvideo, or #dance to create attention for popular descriptive categories (Highfield & Leaver 2015), or #charli or #charlidamelio to relate them to the famous initiator on TikTok. Most frequently, videos included #fyp, #foryou, or #foryoupage, which is a common strategy among TikTok users. On TikTok, the "For You" page is the personalized recommendation feed users first see when opening the app. Therefore, by using these hashtags, users assume to get featured on others "For You" page, though there is no algorithmic or any proof that it works[25]. The same strategy was found for some users adding #xyzbca, which is a nonsense hashtag with no meaning that is supposed to trick algorithms into finding the video[26]. Among the variety of hashtags, #pgpartner, #dancechallenge or #distantdancechallenge, and #stayathome or #quarantine were also used frequently, referring to either the challenge sponsor, as done in Charli's video, the existence of the challenge itself, or the social situation of social distancing during the time of this challenge. Few users also included hashtags to other at the time ongoing challenges or initiatives, such as #earthhour[27], a movement to support the environment by turning off your light on March 28, 2020, or #thesongofus[28], a viral campaign for a music video to be made from TikTok videos.

---

[25] https://www.vice.com/en_us/article/xwezwj/how-does-tiktoks-for-you-page-work-users-have-some-wild-theories [31.3.2020].
[26] https://stayhipp.com/glossary/what-is-xyzbca/ [31.3.2020].

More interestingly, few users used hashtags that seem unrelated to the #distancedance challenge, Charli D'Amelio or the COVID-19 situation. Most noticeably, three users, one male and two female teenager(s), added #hairtutorials and #onlineclass to their videos; however, their TikTok profiles revealed none of them actually has any videos that could be considered a hair tutorial or an online class of some kind, rather they mostly do dance related video performances. This allows to conclude, that users are adding popular hashtags – #hairtutorial has 2.3 billion views[29], #onlineclass has 2 billion views[30] – to generate greater visibility and attention for their video in general, but as well within the already highly popular #distancedance and its sub-community.

Other single uses of hashtags were apparently related to users' biographies, for example, their profession (#nyhygenist), their family status (#singleparents), their location (#tiktokcanada), or their ethnicity (#mixedgirl). A number of used hashtag need more context to understand, for example, a male person used #calvinkleinjeans without either wearing them or being in any observable way connected the product or the company. The sample also showed some misspelled hashtags, such as #blackhirlmagic (instead of girl) or #distansdance (instead of distant).

---

27 https://www.tiktok.com/tag/earthhour?lang=en [31.3.2020].
28 https://www.tiktok.com/tag/TheSongOfUs [31.3.2020].
29 https://www.tiktok.com/amp/tag/hairtutorial?lang=en [2.4.2020].
30 https://www.tiktok.com/tag/onlineclass?lang=en [2.4.2020].

- *<u>Visual Elements and Gestures as Indicators of Performance Practices</u>*

Qualitative content analysis of videos allows to identify phenomena that indicate underlying practices and strategies of users creating videos of themselves performing the #distancedance. In the context of social practices, everyday self-presentation in dance video creation, and social media participation, all performances in the sample illustrate users' individual habitual, bodily, mimic, and gestural characteristics as unintentional "signs given off" in contrast to planned actions and expressions as intentional "signs given" (Goffman 2002 [1959]). The majority of users in the sample performed all moves of the #distancedance in the right order, form, and speed. According to the idea of participating in TikTok dance challenges, this means users put enough practice and effort into learning to perform the dance and subsequently shared a video of them succeeding the challenge. Especially videos of two or three performers suggest that previous time and effort was needed to learn and coordinate the dance moves for a group performance.

However, a significant number of participants did different dances, performed only some moves of the #distancedance, or performed out of sync. This shows that TikTok dance challenges do not necessarily have to be understood as competitions but as well as casual and fun activities, like making videos and participating in social video communities. In addition, the specific charitable context of the #distancedance challenge serves as increased motivation to participate – even if one is not capable of actually doing the complete or correct dance.

In addition to well rehearsed and performed dances, some videos also indicated well prepared settings, such as professional lighting, and outfits, for example matching clothes in a group performance. This suggests a certain amount of planning put into the video creation. In contrast, other video examples include unplanned elements, such as family members in the background, or one of the performers accidentally hitting another one in the head. Users could have decided to share other video versions instead, which implies various strategies for participating in a dance challenge. For example, users could simply decide to upload the first version they finally get the moves right and not care about distractions; the motivation for users to join the challenge could generally be greater than their ambition to get the dance right; users could have a low barrier for reviewing and evaluating their performance compared to other performances in the challenge etc.

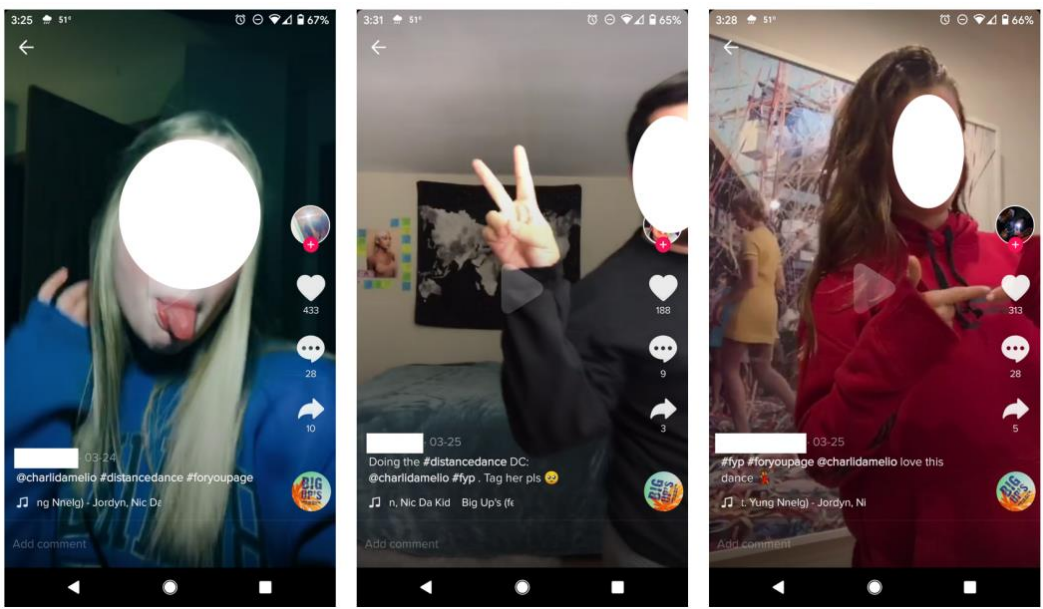

Fig. 7 Examples of additional individual performance ending gestures in the sampled videos
(left to right: sticking tongue out, V sign, imitating "shy emoji" with fingers)

The #distancedance, as well as other dance challenges, is not intended for individual interpretation. It is a fixed set of moves that should be performed in the originally demonstrated way. A useful way to identify performance practices is to look at initiating and ending performative elements at the start and end of a performative act (Schechner 2017). Only two performances included initiating acts, one user can be seen stepping away from the phone camera after starting the recording, another user steps into the frame; however, both users perform moves different from the #distancedance.

A significant number of performances include performance ending or closing elements. Most common is walking off frame to mark the end of the performance, hence, it is part of Charli's original performance copied by the majority of users. A significant number of users add gestures as performative elements at the end of the dance performance. As Rettberg (2017) shows for musical.ly, hand signs can be understood as codified constituents in non-verbal video app communication. These short additional elements may be planned with care given the 15 second limit of TikTok videos and intended to add individual and exceptional elements to the dance imitation. Yet, the sampled videos reveal a quiet homogenous use of additional gestures that as well relate to current popular phenomena in TikTok and social media communication. In most cases, users made the "V sign" while walking off frame. While in Western culture being either an insulting gesture or a symbol for peace or victory, in East Asian culture the "V sign" is closely linked to K-Pop culture and the "aegyo" trend ("behaving cute") (Puzar & Hong 2018) and used when posing in party pictures or selfies[31]. This suggests users adapt the "V sign" as a cross-cultural symbol that is strongly linked to East Asian originated social media culture, such as TikTok, respectively Douyin. Some users imitate the "two fingers touching" emoji (👉👈) which, especially on TikTok, means "being shy" or "hesitate"[32]. When used at the end of a video, users might do this gesture to communicate uncertainty about their performance or or being uncomfortable with it.

- *Paratextual Indicators of Performance Practices*

---

[31] https://time.com/2980357/asia-photos-peace-sign-v-janet-lynn-konica-jun-inoue [2.4.2020].
[32] https://knowyourmeme.com/memes/%F0%9F%91%89%F0%9F%91%88-two-fingers-touching [2.4.2020].

Some users added textual information to their videos or captions that point to performance related production practices. Most commonly, these referred to time and effort needed to perform the dance correctly, for example "do i have school tomorrow? yes. did i stay up to learn this? yes :)" or "OmO it took me almost 30 minutes to practice this". Textual information reveals the greater cause of this challenge as motivation to participate ("Did the #distancedance by @charlidamelio it's for a great cause!") or even to do dance videos at all ("First tiktok dance ever. Doing it for P&G Feeding America and Matthew 25"). Captions as well imply participation in the #distantdance triggered by peer community belonging ("only bc maddy made me") or by the emergence of the TikTok phenomenon as such ("had to hit this new dance quick before bed"). Captions also indicate that users rehearsed and reviewed their performances before posting a video ("best of my many attempts", "The out outcome of 2.524 tries…"). Some users used captions to comment and reflect on their dance performance ("ceo of forgetting the dance"), learning the moves ("Wow I dunno this part LOL"), or their interpretation of it ("i went kind hard whoops").

Videos showing unplanned elements ("Featuring my pro dancing dad") or performance mishaps ("I hit her at the end") used captions as explanation and justification for including these actions in the video. This might indicate that the needed rehearsing effort outweighs these deviances in the decision to (not) share this dance version. Captions may also give information about intentions in planning the dance performance, such as choosing a certain outfit, for example, motivated by the COVID-19 quarantine situation ("Can staying inside & dressing like its vacation become a trend?").

**Conclusion**

The results from qualitative content analysis of a sample of TikTok videos submitted to the #distancedance challenge show that, in a first step, strategies and performance practices of users participating in the challenge. The sample reveals that #distancedance challenge videos are mainly performed by single white female teenagers wearing casual outfits in their bedrooms.

In case of the #distancedance, the video captions illustrate that many users were motivated to participate by the greater cause of the challenge. This seems to be a specific aspect of this challenge; however, some users revealed they were inspired by friends who were doing the challenge or by very likely following Charli D'Amelio. Through the TikTok "duet" videos, the sample also showed that users consume videos of challenge participants other than Charli D'Amelio and use others' #distancedance performance as background for their own performance in a "duet". This indicates a way of how users become aware of TikTok dance challenges. In video captions, users shared their experiences about learning and performing the dance moves, mostly by referring to the time needed to master the dance. This points to the effort users are willing to put into participating in challenges that involve performance activities and self-presentation.

Some strategies of planning surrounding the video performance can be detected in the sample. For example, users prepared settings, chose various lighting, and indicated in captions that they specifically dressed up for recording their dance performance. However, this is contrasted by the majority of rather unplanned or spontaneous looking videos. This indicates most videos might be part of a series of user attempts to master the dance challenge resulting in posting the first successful video performance to TikTok. Videos that show a prepared setting and outfit could mean that users rehearsed their dance performance to a point where they are confident to master it in that recording.

On TikTok, the presentation of the self and production practices are largely connected to performing to short musical or sound snippets. While dance challenges are predefined moves for participants to copy, the video sample also shows that users add gestures to their performances that derive from video-based social media communication. Participants who are routine users of TikTok and similar social video services might have internalized these gestures as common signals to include in their video performances. By adding these signals, users show their knowledge of using them as part of an online community while at the same time manifesting their belongingness to the community.

This exemplary study provides first insights based on video content analysis as product analysis. There are obvious limitations regarding the analysis and reconstruction of participants' consumption strategies, motivations, practices of rehearsing and reviewing recorded performances, and aspects of follow-up consumption or monitoring of their own and other participants' videos.

Based on these first findings, this calls for additional investigation and validation through various ethnographic approaches to amateur production based on media production studies (Banks, Conor & Mayer 2015, Mayer, Banks & Caldwell 2009). Most suitable, qualitative interviews with participants and video observations of their production practices could help to document all stages of a video creation process, that is, consuming, planning, rehearsing, filming, reviewing, editing, sharing, monitoring, and repeating as TikTok video creation practices. This research is also not limited to TikTok but concerns all short video apps, for example, Lasso, Likee, or Vigo Video, and similar services, such as Instagram videos (Boomerang) and stories, or Snapchat stories. Further research is planned using a combined methodological approach of production and product analysis.

**References**

Ahlse, J., Nilsson, F., & Sandström, N. (2020). It's time to TikTok: Exploring Generation Z's motivations to participate in #Challenges. Bachelor Thesis. Jönköping University. URL: https://www.diva-portal.org/smash/get/diva2:1434091/FULLTEXT01.pdf [2.7.2020].

Altheide, D. L., & Schneider, C. J. (2012). *Qualitative media analysis* (Vol. 38). Sage Publications.

Anderson, K. E. (2020). Getting acquainted with social networks and apps: it is time to talk about TikTok. *Library Hi Tech News*.

Banks, M., Conor, B., & Mayer, V. (Eds.). (2015). *Production studies, the Sequel!: cultural studies of global media industries*. Routledge.

Bergman, S. M., Fearrington, M. E., Davenport, S. W., & Bergman, J. Z. (2011). Millennials, narcissism, and social networking: What narcissists do on social networking sites and why. *Personality and Individual Differences*, *50*(5), 706-711.

Bordwell, D., Thompson, K., & Smith, J. (2020). *Film art: An introduction* (Vol. 12). New York: McGraw-Hill.

Bresnick, E. (2019). Intensified Play: Cinematic study of TikTok mobile app. URL: www. researchgate. net/publication/335570557_Intensified_Play_Cinematic_study_of_TikTok _mobile_app [2.4.2020].

Burgess, J. (2014). 'All your chocolate rain are belong to us'?: Viral video, YouTube and the dynamics of participatory culture. *Art in the global present [CSR Books, Number1]:*, 86-96.

Burgess, J., & Green, J. (2018). *YouTube: Online video and participatory culture*. John Wiley & Sons.

Burns, A. (2015). Self (ie)-Discipline: Social Regulation as Enacted Through the Discussion of Photographic Practice. *International Journal of Communication*, *9*, 18.

Cayari, C. (2011). The YouTube Effect: How YouTube Has Provided New Ways to Consume, Create, and Share Music. *International Journal of Education & the Arts*, *12*(6), n6.

Chen, Z., He, Q., Mao, Z., Chung, H. M., & Maharjan, S. (2019, May). A study on the characteristics of douyin short videos and implications for edge caching. In *Proceedings of the ACM Turing Celebration Conference-China* (pp. 1-6).

Evans, C., Hackney, R., Rauniar, R., Rawski, G., Yang, J., & Johnson, B. (2014). Technology acceptance model (TAM) and social media usage: an empirical study on Facebook. *Journal of Enterprise Information Management*. *27*(1), 6-30.

Flick, U. (2014). *An introduction to qualitative research*. Sage.

Genette, G. (1997). *Paratexts: Thresholds of interpretation* (Vol. 20). Cambridge University Press.

Giddens, A. (2011 [1984]). *The constitution of society: Outline of the theory of structuration*. Univ of California Press.

Goffman, E. (2002 [1959]). The presentation of self in everyday life. 1959. *Garden City, NY*, *259*.

Guerrero-Pico, M., Masanet, M. J., & Scolari, C. A. (2019). Toward a typology of young produsers: Teenagers' transmedia skills, media production, and narrative and aesthetic appreciation. *New Media & Society*, *21*(2), 336-353.

Highfield, T., & Leaver, T. (2015). A methodology for mapping Instagram hashtags. *First Monday*, *20*(1), 1-11.

Huang, C. Y., Chou, C. J., & Lin, P. C. (2010). Involvement theory in constructing bloggers' intention to purchase travel products. *Tourism Management*, *31*(4), 513-526.

Ito, M., Baumer, S., Bittanti, M., Cody, R., Stephenson, B. H., Horst, H. A., ... & Perkel, D. (2019). *Hanging out, messing around, and geeking out: Kids living and learning with new media*. MIT press.

Jenkins, H. (2009). *Confronting the challenges of participatory culture: Media education for the 21st century*. Mit Press.

Khattab, M., (2019). Synching and performing : body (re)- presentation in the short video app TikTok. WiderScreen 21(1– 2). http://widerscreen.fi/numerot/2019-1-2/synching-andperforming-
body-re-presentation-in-the-short-video-apptiktok/ [4.4.2020].

Lange, P. G. (2014). *Kids on YouTube: Technical identities and digital literacies*. Left Coast Press.

Leung, L. (2009). User-generated content on the internet: an examination of gratifications, civic engagement and psychological empowerment. *New media & society*, *11*(8), 1327-1347.

Lu, X., & Lu, Z. (2019, July). Fifteen Seconds of Fame: A Qualitative Study of Douyin, A Short Video Sharing Mobile Application in China. In *International Conference on Human-Computer Interaction* (pp. 233-244). Springer, Cham.

Manovich, L. (2009). The practice of everyday (media) life: From mass consumption to mass cultural production?. *Critical Inquiry*, *35*(2), 319-331.

Mayer, V., Banks, M. J., & Caldwell, J. T. (Eds.). (2009). *Production studies: Cultural studies of media industries*. Routledge.

McGloin, R., & Oeldorf-Hirsch, A. (2018). Challenge accepted! Evaluating the personality and social network characteristics of individuals who participated in the ALS Ice Bucket Challenge. *The Journal of Social Media in Society*, *7*(1), 443-455.

McRoberts, S., Ma, H., Hall, A., & Yarosh, S. (2017, May). Share first, save later: Performance of self through Snapchat stories. In *Proceedings of the 2017 CHI Conference on Human Factors in Computing Systems* (pp. 6902-6911).

Phua, J., Jin, S. V., & Kim, J. J. (2017). Uses and gratifications of social networking sites for bridging and bonding social capital: A comparison of Facebook, Twitter, Instagram, and Snapchat. *Computers in human behavior*, *72*, 115-122.

Puzar, A., & Hong, Y. (2018). Korean Cuties: Understanding Performed Winsomeness (Aegyo) in South Korea. *The Asia Pacific Journal of Anthropology*, *19*(4), 333-349.

Rettberg, J. W. (2017). Hand signs for lip-syncing: the emergence of a gestural language on musical.ly as a video-based equivalent to emoji. *Social Media+Society*, *3*(4), 2056305117735751.

Schatzki, T. R. (1996). *Social practices: A Wittgensteinian approach to human activity and the social*. Cambridge University Press.

Schechner, R. (2017). *Performance studies: An introduction*. Routledge.

Scolari, C. A., & Fraticelli, D. (2019). The case of the top Spanish YouTubers: Emerging media subjects and discourse practices in the new media ecology. *Convergence*, *25*(3), 496-515.

Simonsen, T. M. (2014). The functionality of paratexts on YouTube. In *Examining paratextual theory and its applications in digital culture* (pp. 209-234). IGI global.

Vandersmissen, B., Godin, F., Tomar, A., De Neve, W., & Van de Walle, R. (2014). The rise of mobile and social short-form video: an in-depth measurement study of vine. In *Workshop on Social Multimedia and Storytelling* (Vol. 1198, pp. 1-10).

Vaterlaus, J. M., Barnett, K., Roche, C., & Young, J. A. (2016). "Snapchat is more personal": An exploratory study on Snapchat behaviors and young adult interpersonal relationships. *Computers in Human Behavior*, *62*, 594-601.

Yarosh, S., Bonsignore, E., McRoberts, S., & Peyton, T. (2016, February). YouthTube: Youth video authorship on YouTube and Vine. In *Proceedings of the 19th ACM Conference on Computer-Supported Cooperative Work & Social Computing* (pp. 1423-1437).

Yau, J. C., & Reich, S. M. (2019). "It's Just a Lot of Work": Adolescents' Self-Presentation Norms and Practices on Facebook and Instagram. *Journal of research on adolescence*, *29*(1), 196-209.

Zhang, L., Wang, F., & Liu, J. (2014, March). Understand instant video clip sharing on mobile platforms: Twitter's vine as a case study. In *Proceedings of Network and Operating System Support on Digital Audio and Video Workshop* (pp. 85-90).

Zhang, X., Wu, Y., & Liu, S. (2019). Exploring short-form video application addiction: Socio-technical and attachment perspectives. *Telematics and Informatics*, *42*, 101243.